\documentstyle[amssymb,aps,epsfig]{revtex}

\begin{document}
\title{Role of dynamic Jahn-Teller distortions in Na$_{2}$C$_{60}$ and Na$_{2}$CsC$%
_{60}$ studied by NMR}
\date{\today}
\author{ V. Brouet$^{1}$, H. Alloul$^{1}$, T.N.\ Le$^{2}$, S. Garaj$^{2}$ and L. Forr{%
\'{o}}$^{2}$}
\address{$^{1}$Laboratoire de Physique des Solides, Universit\'{e} Paris-Sud, B\^{a}t
510, 91 405 Orsay (France)\\
$^{2}$Laboratoire de Physique des Solides Semicristallins,\\
IGA-D\'{e}partement de Physique, Ecole Polytechnique F\'{e}d\'{e}rale de
Lausanne, 1015 Lausanne (Switzerland)}

\twocolumn[\hsize\textwidth\columnwidth\hsize\csname@twocolumnfalse\endcsname

\maketitle

\begin{abstract}
Through $^{13}$C NMR spin lattice relaxation (T$_{1}$) measurements in cubic
Na$_{2}$C$_{60},$ we detect a gap in its electronic excitations, similar to
that observed in tetragonal A$_{4}$C$_{60}$. This establishes that
Jahn-Teller distortions (JTD) and strong electronic correlations must be
considered to understand the behaviour of even electron systems, regardless
of the structure. Furthermore, in metallic Na$_{2}$CsC$_{60}$, a similar
contribution to T$_{1}$ is also detected for $^{13}$C and $^{133}$Cs NMR,
implying the occurence of excitations typical of JT distorted C$_{60}^{2-}$
(or equivalently C$_{60}^{4-}$). This supports the idea that dynamic JTD can
induce attractive electronic interactions in odd electron systems.
\end{abstract}

]

It has been known, almost since the discovery of alkali doped fullerenes,
that A$_{4}$C$_{60}$ is insulating, while A$_{3}$C$_{60}$ is metallic and
superconducting \cite{Haddon}. This contrasts with expectations for a rigid
band filling model, as the C$_{60}$ lowest unoccupied molecular t$_{1u}$
level should form a triply degenerate band.\ All A$_{n}$C$_{60}$ with n%
\mbox{$<$}%
6 should either be metals if the strength of the Coulomb repulsion U is
small compared to the band width W or Mott insulators if U is larger. In
fullerides, U/W is close to the critical ratio where a metal-insulator
transition is expected \cite{GunnarssonPRB96}, but the observation of both
metals and insulators within this family is puzzling. 

An attractive explanation \cite{Heritier,Manini,Gunnarsson} is that the
presence of Jahn-Teller distortions (JTD) could create {\it effective
electronic interactions }which would modulate U, so that different compounds
could be on different sides of the metal-insulator transition. More
precisely, the computation of U$_{eff}$ = E(n+1)+E(n-1)-2 E(n), where n is
the number of electrons per C$_{60},\ $supports this idea when the larger
gain of electronic energy associated with the JTD of an evenly charged C$%
_{60}$ ball is taken into account \cite{Heritier,Manini,Gunnarsson}. It
follows that in even electron systems JT effects {\it add} to the Coulomb
repulsion to localize electrons, whereas for odd electrons systems they {\it %
oppose }the Coulomb repulsion and favor delocalization. Hence, A$_{2n}$C$%
_{60}$ could be insulating whereas A$_{2n+1}$C$_{60}$ could become metallic
despite the strong electronic correlations. To give some experimental
support to this idea, a large amount of work has been devoted to the search
of JTD in fullerides, but without success up to now. However, the distortion
is expected to be very small and possibly dynamic, which makes it difficult
to detect directly. Alternatively, recent models have suggested that the
different properties between A$_{4}$C$_{60}$ (body centered tetragonal) and A%
$_{3}$C$_{60}$ (face cubic centered) are due to their different structures 
\cite{GunnarssonA4}. 

In order to sort out the relevant parameters for the physics of A$_{n}$C$%
_{60}$, we present here an NMR study of Na$_{2}$C$_{60}$ and Na$_{2}$CsC$%
_{60}$, as they have the same{\it \ cubic} structure but {\it even} and {\it %
odd} stoichiometries. Na$_{2}$C$_{60}$ is the only compound with n=2 known
so far, but experimental studies are still limited and controverted. The
first ESR studies \cite{Petit} concluded that its electronic properties were
identical to K$_{4}$C$_{60}$ but more recently, another ESR investigation
claimed that it was metallic with a metal-insulator transition at 50 K \cite
{Kubozono}. We present the first NMR study of Na$_{2}$C$_{60}$, which allows
to detect singlet-triplet excitations of JTD C$_{60}$ balls, as found in A$%
_{4}$C$_{60}$ \cite{ZimmerK4,Kerkoud,ZimmerPRB95}. This rules out a strong
dependence of the electronic properties on the structure, as one expects
similar properties for n = 2 and 4 due to electron-hole symmetry in the t$%
_{1u}$ band. We then extend our investigation to Na$_{2}$CsC$_{60},$ which
superconducts below $T_{c}=12$ $K$ \cite{Tanigaki}.\ In the metallic phase,
we evidence an anomalous contribution to the $^{13}$C and $^{133}$Cs NMR
spin-lattice relaxation rate $1/T_{1}.$ As it is similar to that found in Na$%
_{2}$C$_{60}$, we assign it to the presence of Jahn-Teller distorted C$%
_{60}^{(2,4)-}.\,$This demonstrates the importance of dynamic JTD in
metallic fullerides also and suggests that the metallic character of A$_{3}$C%
$_{60}$ could be related to an enhanced stability of C$_{60}^{(2,4)-}$%
.\medskip\ 

The Na$_{2}$C$_{60}$ and Na$_{2}$CsC$_{60}$ samples were prepared by
conventional solid-state reaction. Phase purity was checked by X-ray
diffraction. At high $T,$ both compounds have the same structure as other A$%
_{3}$C$_{60}$ systems (face centered cubic structure (fcc) with space group
Fm$\overline{3}$m) but they undergo below room $T$ an orientational ordering
transition like pure C$_{60}$ (the symmetry is reduced to simple cubic (sc)
with Pa$\overline{3}$ space group) \cite{Yldirim,Prassides}.\ NMR
measurements were carried out in a 7~T field and spin-lattice relaxation
measurements were obtained with usual saturation recovery sequences.

\begin{figure}[t]
\epsfig{%
file=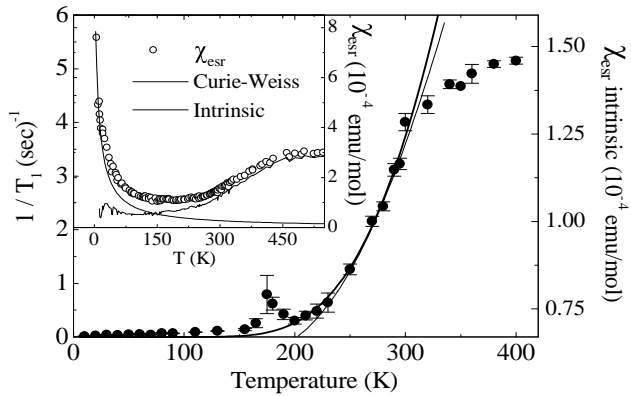,%
figure= Newfig1.eps,%
height=5.5 cm,%
width=8.5 cm,%
angle=0,%
}
\caption{$^{13}$C NMR 1/T$_{1}$ as a function of temperature in Na$_{2}$C$%
_{60}$. The thick line is a fit to an activated law with E$_{g}$=140 meV. The thin line (right scale) is the intrinsic part of $\protect%
\chi _{esr}$. Inset : $\chi {esr}$ in Na$_{2}$C$%
_{60}$ decomposed into an intrinsic part and a Curie-Weiss contribution. }
\label{Na2T1car}
\end{figure}

Figure \ref{Na2T1car} shows that $1/T_{1}$ for $^{13}$C in Na$_{2}$C$_{60}$
increases very steeply and can be modelled by an activated law $%
1/T_{1}\propto exp(-E_{a}/k_{B}T)$ with $E_{a}=140\pm 20$ meV. This
contrasts with the linear $T$ dependence (the Korringa law) expected for
simple metals \cite{hyperfin} and indicates that a gap $E_{a}$ separates the
ground state from the excited states of the system. This is similar to the
insulating A$_{4}$C$_{60}$ systems, where an activated behaviour dominates
the relaxation with $E_{a}\simeq 50-75$ meV \cite
{ZimmerK4,Kerkoud,ZimmerPRB95,convention}. Besides this central fact, more
features are evident :

i)There is a peak at 180 K which is typical of a 

\begin{figure}
\epsfig{%
file=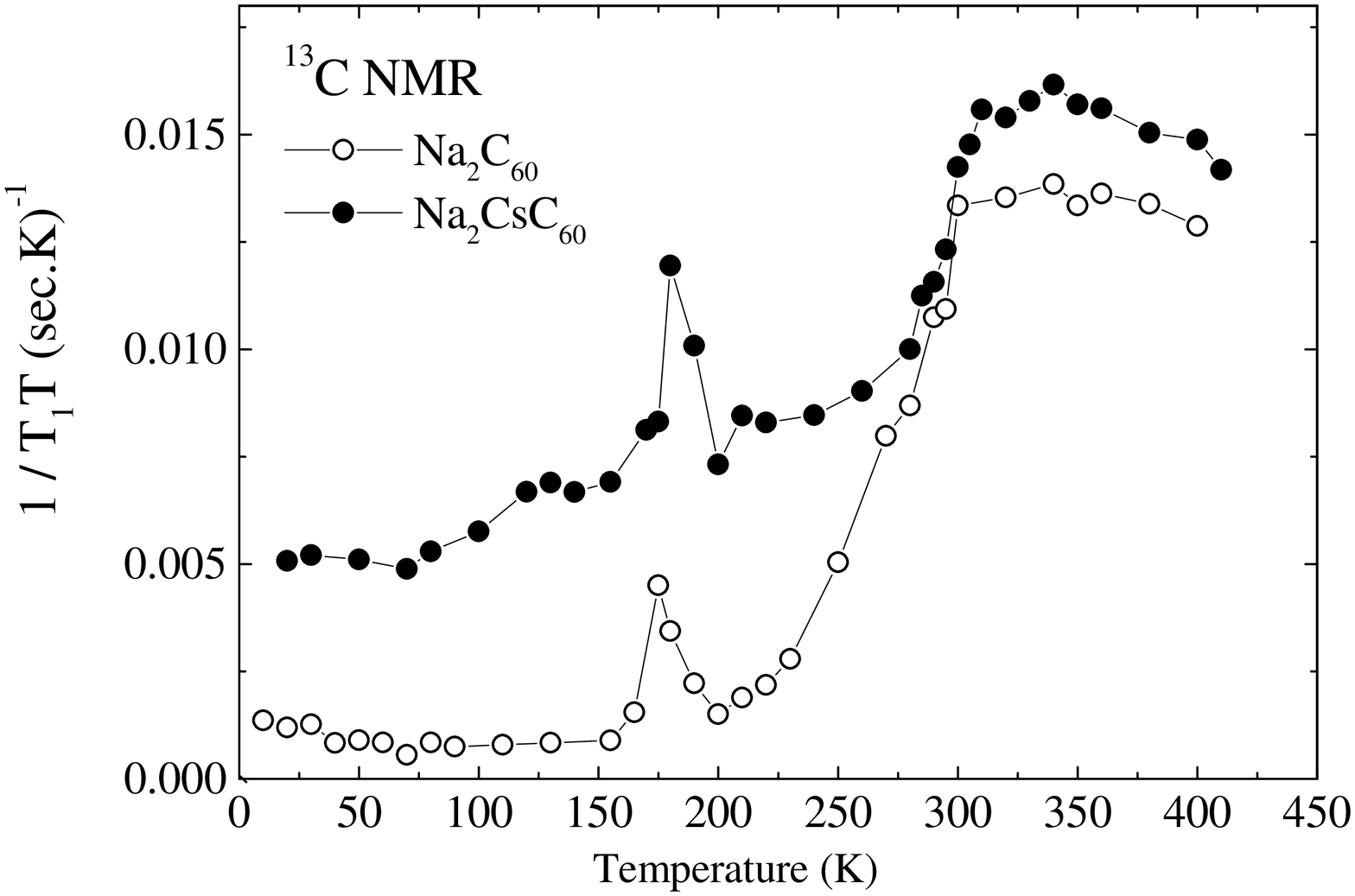,%
figure= CarNaNa2.eps,%
height=4.7 cm,%
width=8 cm,%
angle=0,%
}
\caption{1/T$_{1}$T for $^{13}$C as a function of temperature in Na$_{2}$C$%
_{60}$ and Na$_{2}$CsC$_{60}$. Below 200 K, the recovery curves for the magnetization are not
exponentional and T$_{1}$ is defined as the mean value of a double exponential fit.}
\label{CarNa2Na2Cs}
\end{figure}

contribution due to the slowing down of C$_{60}$ molecular motions \cite{Yoshinari}, as shown in
more details elsewhere \cite{Kirchberg2000}.

ii) At low temperature, $1/T_{1}$ is enhanced with respect to the activated
law. On Figure \ref{CarNa2Na2Cs}, it can be more clearly seen that $%
(T_{1}T)^{-1}$ tends to a constant value. If intrinsic, such a contribution
could indicate a residual metallic character with a small density of states $%
n(E_{f})\simeq 1$~eV$^{-1}$spin$^{-1}$ \cite{hyperfin}. A very similar
behaviour was in fact observed in Rb$_{4}$C$_{60},$ where an additional
relaxation mechanism becomes efficient at low $T$, which was assigned to a
small gap ($\simeq 20$ meV) \cite{Kerkoud}. This gap is easily closed by
applying pressure and a growing linear contribution to the relaxation is
observed with increasing pressure {\it that coexists} with the activated
contribution \cite{Kerkoud}. By analogy, Na$_{2}$C$_{60}$ would be
equivalent to Rb$_{4}$C$_{60}$ under an applied pressure of roughly $1$~kbar.

The relaxation behaviour does not change down to 10~K, implying that there
is no transition to an ordered ground state. In particular, we do not
observe any anomaly at 50 K that might be assigned to a metal-insulator
transition as claimed in ref. \cite{Kubozono}. Hence, the ground state is
non-magnetic, which is confirmed by the absence of magnetic
broadening of the NMR spectra at low T \cite
{Kirchberg2000}. This is again similar to A$_{4}$C$_{60}$.

iii) Above 300K, $1/T_{1}$ saturates, which would only be expected for $%
T\gtrsim E_{a}/k_{B}\approx 1600\,\ K$. As the structural transition from sc
to fcc takes place at 310 K, it is natural to wonder whether there is an
associated change of the electronic properties, for example a smaller gap in
the fcc phase. Data at higher temperatures would be needed to conclude this
unambiguously.

Like 1/T$_{1}$, the ESR susceptibility $\chi _{esr\text{ }}$increases
between 200\ and 300 K \cite{Petit} and a quantitative comparison with NMR
should give us further insight into the properties of the system. Therefore,
we have measured $\chi _{esr\text{ }}$in our sample (see inset of Fig. \ref
{Na2T1car}), with results analogous to those of ref. \cite{Petit}. The
intrinsic part $\chi _{int}\,\,$of $\chi _{esr}$ is difficult to extract
precisely below 250~K as a large Curie contribution is always observed in Na$%
_{2}$C$_{60}$. From a low $T$ fit, we deduce a Curie-Weiss contribution$%
\,\,C/(T+T_{n})\,$with $T_{n}=8$ $K$ corresponding to 2.3 \% impurities per C%
$_{60}.$ As can be seen in Fig \ref{Na2T1car}, the remaining part $\chi
_{int}\,$tends to a constant value of 6.10$^{-5}$~emu/mol at low $T$.
Although direct conductivity measurements will be necessary to conclude
about a possible weak metallicity of this compound, we note that such a
Pauli contribution in $\chi _{int}$ would be consistent with the constant $%
(T_{1}T)^{-1}$ found at low $T$ that we have previously discussed. The $T$
dependent contribution $\chi \,\,$to $\chi _{int}$ can then be compared to $%
1/T_{1}$ (see right scale of Fig. \ref{Na2T1car}). The simplest way to
relate $1/T_{1}$ and $\chi $ is to assume that both are associated with
electronic excitations characterized by a spin correlation function with an
exponential decay time $\tau .$ In the limit $\omega _{e}\tau <<1$, where $%
\omega _{e}$ is the electronic Larmor frequency, one expects \cite{White} :

\begin{equation}
\frac{1}{T_{1}}=\left( \frac{A}{\hbar }\right) ^{2}\frac{\chi }{N_{A}\text{ }%
\mu _{B}^{2}}\text{ }k_{B}T\text{ }\tau 
\end{equation}
where $A$ is the hyperfine coupling and $N_{A}$ the Avogadro number. Fig. 
\ref{Na2T1car} evidences the validity of this scaling between $1/T_{1}$ and $%
\chi .\;$Unfortunately, the limited $T$ range of the experiment does not
allow to probe efficiently the $T$ dependence of $\tau .$ Assuming that it
is constant and using $A~=~4.~10^{-20}$~erg \cite{hyperfin}, we obtain $\tau
\simeq 8.$~$10^{-14}$ sec, which has the same order of magnitude as that
found in Rb$_{4}$C$_{60}$ and is consistent with $\omega _{e}\tau <<1.$

To summarize, we conclude that {\it there are strong similarities between Na}%
$_{2}${\it C}$_{60}${\it \ and A}$_{4}${\it C}$_{60}$. To describe the
weakness of the metallic character of these compounds, models involving a
(dynamic) JTD of the C$_{60}$ molecule that lifts the degeneracy of the t$%
_{1u}$ levels are the most likely, as they naturally yield a non-magnetic
ground state. Indeed, although it could be singlet or triplet, depending on
the nature of the JTD, molecular calculations indicate that the singlet
state has the lowest energy \cite{Manini} for both n=2 and n=4. However, as
Hund's rules favor the triplet state, the two states lie close in energy and
singlet-triplet like excitations would take place when one molecule goes
from one distortion to the other. The 140 meV value of the gap observed in $%
1/T_{1}$ corresponds to the molecular estimate for this ``spin-gap'' \cite
{Manini}. We can then attribute the relaxation to a coupling between the NMR
nuclei and the thermally populated triplet states, which have a spin
lifetime $\tau $. While ref. \cite{Manini} predicts the same spin-gap for C$%
_{60}^{2-}\,$and C$_{60}^{4-}$, the experimental difference between Na$_{2}$C%
$_{60}$ and A$_{4}$C$_{60}\,$could be due to their different structures, as
suggested by the saturation in $1/T_{1}$ observed here at the structural
transition. Alternatively, electron-hole symmetry in the t$_{1u}$ level
might be broken when excitations to higher electronic levels (t$_{1g}$) are
taken into account \cite{OBrien}.

The fact that molecular properties on an energy scale of 140~meV are not
smeared out by the formation in the solid of bands of typically 500 meV width requires strong electronic correlations. This was recognized by 

Fabrizio {\it et al.}\cite{Fabrizio}, who qualified these systems as ``Mott
Jahn-Teller insulators''. They emphasized that the splitting between $t_{1u}$
levels induced by the JT distortion, estimated to be 500 meV (and observed
experimentally as an ``optical gap'' in A$_{4}$C$_{60}$ \cite{Fink,Iwasa})
is too small to lead to a band insulator. Electronic correlations increase
the average time spent by one electron on a C$_{60}$ ball, so that
``molecular physics'', such as the JT distortions, can take place even
before complete localization. This might be the case in Na$_{2}$C$_{60}$,
where a residual metallic character is suggested by our low $T$ data.\

We now turn to the study of Na$_{2}$CsC$_{60}$, for which Fig. \ref
{CarNa2Na2Cs} shows $(T_{1}T)^{-1}$ compared with Na$_{2}$C$_{60}$. Below
150K, $(T_{1}T)^{-1}$ in Na$_{2}$CsC$_{60}$ is dominated by a $T$
independent contribution, in agreement with its metallic character, which is
nearly suppressed in Na$_{2}$C$_{60}.\;$But at higher $T$, $(T_{1}T)^{-1}$
departs from the metallic behaviour and surprisingly, its overall behaviour
is very similar to that of Na$_{2}$C$_{60}$.\ We\ want to argue here that 
{\it this is not accidental} but reveals a similar relaxation mechanism in
the two compounds. We will restrict our discussion to the $sc$ phase of Na$%
_{2}$CsC$_{60}$ ($T$~%
\mbox{$<$}%
~300~K), since it was recently suggested that the $fcc$ phase might be
insulating \cite{Cegar}.

\begin{figure}[tbp]
\epsfig{%
file=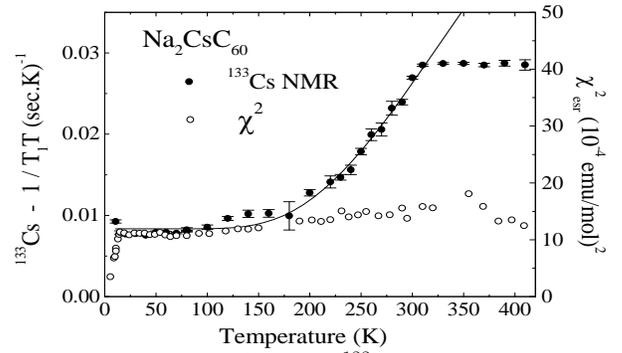,%
figure= CsT1.eps,%
height=4.7 cm,%
width=8 cm,%
angle=0,%
}

\caption{Left scale : 1/T$_{1}$T for $^{133}$Cs as a function of temperature
in Na$_{2}$CsC$_{60}$. The recovery curves are exponential through the full
T range. As shown by the line, the relaxation can be divided into
1/T$_{1}$T=A+B/T*exp(-E$_{g}$/T) with E$_{g}$=110 meV. Right
scale : $\protect\chi ^{2}$ as a function of T.}
\label{CsInvT1}
\end{figure}

The strong deviation from the Korringa law in Na$_{2}$CsC$_{60}$ had already
been observed previously \cite{Maniwa,Saito}. It was attributed to an
increase of the density of states associated to the lattice expansion plus a
peak due to molecular motions around 300 K. We present here new experimental
data to refute this hypothesis. First, to avoid completely a contribution of
the C$_{60}$ molecular motions to the relaxation, we have performed
measurements on $^{133}$Cs which is not coupled to these motions. As can be
seen on Fig. \ref{CsInvT1}, $(T_{1}T)^{-1}$ for $^{133}$Cs does not exhibit
any molecular motion peak around 180 K. Such a peak, already discussed for Na%
$_{2}$C$_{60},$ is also present in the $^{13}$C relaxation of Na$_{2}$CsC$%
_{60}$ but was missed in the study of ref. \cite{Maniwa,Saito}. On the other
hand, as for $^{13}$C, $(T_{1}T)^{-1}$ for $^{133}$Cs does deviate above
150~K\ from the Korringa law, ensuring that this deviation is not due to
molecular motions but to {\it electronic excitations}. Second, to determine
an eventual $n(E_{f})$-related increase of $(T_{1}T)^{-1}$,

 we have measured 
$\chi _{esr}$ in our sample. It does not follow a simple Pauli law, but
increases slightly above 100$~$K. The increase of $(T_{1}T)^{-1}$ should
scale with $\chi ^{2}$, if both were related to a variation of $n(E_{f})$ 
\cite{hyperfin}$,$ but Fig. \ref{CsInvT1} shows that $(T_{1}T)^{-1}$
increases much more steeply, already at 200 K, well into the $sc$ phase.
Hence, {\it the increase of }$(T_{1}T)^{-1}${\it \ in }Na$_{2}$CsC$_{60}$%
{\it \ is related to an additional relaxation channel. }It can be fitted by
an activated law with $E_{a}=110\pm 5$~meV, as sketched on Fig. \ref{CsInvT1}%
.

It is then natural to propose that this additional relaxation mechanism is
similar to that proposed for Na$_{2}$C$_{60}$. For a JT distorted C$%
_{60}^{3-}$, there are no possibilities for singlet-triplet transitions
directly similar to the ones of a C$_{60}^{2-}$. Therefore we believe that
the similarity between Na$_{2}$C$_{60}$ and Na$_{2}$CsC$_{60}$ comes
directly from the presence of C$_{60}^{2-}$ in Na$_{2}$CsC$_{60}$, or C$%
_{60}^{4-}$ which are equally likely to be formed when electrons jump from
ball to ball in the metal \cite{comment}. As the gap observed by NMR in Na$%
_{2}$C$_{60}$ is related to {\it individual} excitations of C$_{60}^{2-}$
and not to a band gap, similar excitations\ might occur in Na$_{2}$CsC$_{60}$
as well, {\it if C}$_{60}^{(2,4)-}${\it \ exist within the metal for times }$%
\tau _{pair}${\it \ sufficiently long compared to the spin lifetime }$\tau
\,\,${\it of an excited state}. This does not imply a static charge
separation in Na$_{2}$CsC$_{60}$, which is ruled out by the $^{23}$Na NMR
spectra (not shown) which display one narrow line shifted from the position
of Na$_{2}$C$_{60}$.

Using Eq. 1, $\tau $ can be extracted from $(T_{1}T)^{-1}$ if the
singlet-triplet component in $\chi $ is known. We assume a similar $T$
dependence as that found in Na$_{2}$C$_{60}$, which scales as $\chi
~=~\alpha T~^{-1}exp~(-E_{a}/T)$ in the experimental $T$ range and the gap
value $E_{a}=110$ meV determined accurately from $^{133}$Cs NMR. Such a
contribution to the susceptibility can only be smaller than the increase of $%
\chi _{esr}$, which implies $\alpha \lesssim 2\ $and yields $\tau \simeq
10^{-14}\sec $. On the other hand, the time $\tau _{res}$ spent by one
electron in the vicinity of one C$_{60}$, which is a lower limit for $\tau
_{pair}$, can be estimated by $\tau _{res}\simeq \hbar n(E_{f})\simeq
6.10^{-15}$~sec in Na$_{2}$CsC$_{60},$ which is only slightly shorter than $%
\tau $. Electronic correlations, Jahn-Teller effects, as well as scattering
on a C$_{60}$ ball, would all increase $\tau _{res}$ compared to this simple
estimate, so that we are likely in the limit $\tau _{pair}>\tau $\ where the
C$_{60}^{(2,4)-}$\ can contribute to the NMR\ relaxation.

Both the gap value and $\tau $ are reduced compared with Na$_{2}$C$_{60}$.
The same trend was observed in Rb$_{4}$C$_{60}$, where $\tau $ and the gap
decrease with increasing pressure, {\it i.e.} increasing density of states 
\cite{Kerkoud}. This suggests a direct relation between\ $\tau $ and $%
n(E_{f})$, which raises the question of the origin of the relaxation time $%
\tau $ for triplet states. One could expect such a trend if the triplet
states are relaxed by conduction electrons, although the triplet state is
formed by ``potential'' conduction electrons, making it difficult to
distinguish two different spin species. This would also predict that for
higher density of states like in K$_{3}$C$_{60}$ and Rb$_{3}$C$_{60},$ the
contribution of triplet states to the relaxation almost disappear. We note
that although a ``better'' Korringa law is observed in these compounds \cite
{Tycko}, a 20\%-30\% increase of $(T_{1}T)^{-1}$ was still noticed between
100 K and 300 K. It could in fact have a similar origin, so that the
presence of C$_{60}^{(2,4)-}$ might be a common feature of metallic A$_{3}$C$%
_{60}$ compounds. \medskip 

In conclusion, we have shown that Na$_{2}$C$_{60}$ has a non-magnetic ground
state and that its low energy electronic excitations are characterized by a
140 meV spin-gap. This is very similar to A$_{4}$C$_{60}$ systems and
supports the ``Mott JT scenario'' \cite{Fabrizio} to describe fullerides
with 2 or 4 electrons per C$_{60}$.\ Furthermore, we evidence very similar
electronic excitations in Na$_{2}$CsC$_{60}$, coexisting with typically
metallic ones. In both cases, we assign the spin-gap to singlet-triplet
excitations between two JT distortions of C$_{60}^{2n-}$ balls (n = 1 or 2).
Because the triplet state provides a very efficient relaxation mechanism for
NMR, we could indirectly detect here for the first time the presence of
dynamic JTD in a superconducting fulleride. Our study also implies that Na$%
_{2}$CsC$_{60}$ undergoes rapid charge fluctuations (on a time scale of 10$%
^{-14}$ sec) which create preferentially C$_{60}^{(2,4)-}$. We suggest that
this can be rationalized as a consequence of the JTD which stabilizes evenly
charged C$_{60}$.\ 

A first indication of the role played by JTD in the electronic properties of
fullerides was the observation in the metallic cubic quenched CsC$_{60}$ of%
{\it \ localized C}$_{60}^{2-}$ \cite{BrouetPRL99}. Although this large
difference of charge lifetime remains to be understood, these behaviours
suggest that the key feature behind the physics of cubic fullerides is due
to an interplay between strong electronic interaction and ``JT mediated
electronic interactions''. Electronic correlations are essential for
molecular excitations, such as the JTD, to exist in the solid. JT mediated
electronic interactions play in turn a crucial role in determining the
insulating or metallic character of a given compound, by inducing repulsive
interactions in even electron systems and attractive interactions in odd
electron systems, in order to promote the existence of C$_{60}^{2n-}$. The
recent discovery of a way of doping continuously the C$_{60}$ t$_{1u}$ level
through a field-effect device might open a new path for checking this
original behaviour more systematically \cite{Batlogg}.\medskip 

We thank N. Manini and E. Tosatti for stimulating discussions and we
acknowledge the financial support from the TMR program of the European
Commission (``FULPROP'' ERBRMRXVT970155).


\begin{references}
\bibitem{hyperfin}  In a metal, one expects (T$_{1}$T)$^{-1}$ = ($\pi k_{b}/$%
h\hskip-.2em\llap{\protect\rule[1.1ex]{.325em}{.1ex}}\hskip.2em%
)$A^{2}$n(E$_{f}$)$^{2}$. From the Korringa value at low T in Na$_{2}$CsC$%
_{60},$ we deduce an hyperfine coupling $A$ = 4 10$^{-20}$ erg, which lies
with other estimates of $A$ in fullerides within a factor 4.
\bibitem{convention}  We use here E$_{a}=E_{g}$ instead of E$_{a}=E_{g}/2$,
as done in \cite{ZimmerK4,Kerkoud,ZimmerPRB95} by reference to a
semiconducting model.
\bibitem{comment}  For simplicity, we assume here that C$_{60}^{2-}$ and C$%
_{60}^{4-}$ behave similarly, although it could be discussed following ref. 
\cite{OBrien}
\end{references}
\end{document}